\documentclass[twocolumn]{aastex701}

\usepackage{lineno}
\usepackage{amsmath}
\usepackage{cancel}
\usepackage{empheq}
\usepackage{float}

\renewcommand{\vec}[1]{\boldsymbol{#1}}

\newcommand{\be}{\begin{align}}
\newcommand{\ee}{\end{align}}

\shorttitle{Spin and Obliquity Distributions of Low-mass Planets}
\shortauthors{Wang, Li and Lai}

\graphicspath{{./}{figures/}}

\begin{document}

\title{Spin and Obliquity Distributions of Low-mass Planets Shaped by Dynamical Instability}

%\correspondingauthor{Dieran Wang}

\author[0009-0000-3288-4179]{Dieran Wang}
\altaffiliation{dieran.wang@sjtu.edu.cn}
\affiliation{Tsung-Dao Lee Institute, Shanghai Jiao Tong University, 1 Lisuo Road, Shanghai, 201210, China}
\email{dieran.wang@sjtu.edu.cn}

\author[0000-0001-5550-7421]{Jiaru Li}
\altaffiliation{jiaru.li@northwestern.edu}
\affiliation{Center for Interdisciplinary Exploration and Research in Astrophysics (CIERA), Northwestern University, 1800 Sherman Ave, Evanston, IL 60201, USA}
\email{jiaru.li@northwestern.edu}

\author[0000-0002-1934-6250]{Dong Lai}
\altaffiliation{donglai@sjtu.edu.cn}
\affiliation{Tsung-Dao Lee Institute, Shanghai Jiao Tong University, 1 Lisuo Road, Shanghai, 201210, China}
\affiliation{Center for Astrophysics and Planetary Science, Department of Astronomy, Cornell University, Ithaca, NY 14853, USA}
\email{donglai@sjtu.edu.cn}

\begin{abstract}
Exoplanetary systems hosting multiple low-mass planets are thought to have experienced dynamical instability, during which planet-planet collisions and mergers occur;
these collisions can impart substantial amount of angular momentum to the merger remnants, changing the obliquities of the resulting planets significantly.
In this work, we carry out a series of $N$-body experiments to investigate the spin magnitude $(|\vec{S}|)$ and obliquity $(\theta_{\rm SL})$ distributions of low-mass exoplanets that have gone through planetary collisions. 
In our fiducial super-Earth (with $m=3M_{\oplus}$, $R=1.3R_{\oplus}$) and mini-Neptune systems (with $m=9M_{\oplus}$, $R=2.5R_{\oplus}$), the collision products follow a nearly uniform distribution in $\cos{\theta_{\rm SL}}$ and the spin-magnitude distribution is approximately linear in $|\vec{S}|$.
Parameter studies and theoretical analysis show that increasing planetary radii or masses, or decreasing  the initial planet-planet mutual inclinations, tend to polarize the obliquity distribution toward alignment or anti-alignment (i.e., excess probability near $\cos{\theta_{\rm SL}}=\pm1$).
Experiments with initially two-planet and three-planet systems produce qualitatively similar outcomes, suggesting that the trends in this study may generalize to systems with higher planetary multiplicities.
\end{abstract}

\section{Introduction}
\label{sec:Introdution}

The obliquity of a planet, defined as the angle $(\theta_{\rm SL})$ between its spin axis and its orbital angular momentum vector, is an important property that can affect both its climate~\citep[e.g.,][]{Williams.1997.Icar, Spiegel.2009.ApJ, Heller.2011.A&A, Armstrong.2014.AsBio}
and dynamical evolution~\citep[e.g.,][]{Millholland.2019.NatAs, Millholland.2020.ApJ, Su.2022.MNRAS}.
Planets in the Solar Systems are known for having a broad of range obliquities, from nearly zero for Mercury and Jupiter, to the extreme $98^{\circ}$ obliquity for Uranus, to the $177^{\circ}$ retrograde spin for Venus.
This diversity provides both insights and puzzles on the formation history of the Solar System~\citep[e.g.,][]{Lissauer.1993.ARA&A}.

For exoplanets, however, obliquities are extremely difficult to measure.
Obliquity measurements have only been reported for less than a handful of planetary mass objects~\citep{Bryan.2020.AJ, Bryan.2021.AJ, Palma-Bifani.2023.A&A, Poon.2024.AJ}.
Hence, for over 5000 exoplanets discovered to date, theoretical understanding of what shapes their spin and obliquity distributions are valuable.

Planetary obliquity can be shaped by a variety of processes, including the giant impacts/collisions with planetesimals~\citep[e.g.,][]{Safronov.1969.Icar, Lissauer.1991.Icar, Dones.1993.Sci, Morbidelli.2012.Icar}, the spin-orbit resonances~\citep[e.g.,][]{Colombo.1966.AJ, Goldreich.1966.AJ, Laskar.1993.Natur, Millholland.2019.ApJ, Su.2020.ApJ,Su.2022.MNRAS.513}, and the tidal interaction between short-period planets and their host stars~\citep[e.g.,][]{Winn.2005.ApJ, Fabrycky.2007.ApJ, Levrard.2007.A&A, Su.2022.MNRAS}.
For exoplanets, planet-planet collisions are especially important.
These catastrophic events can impart substantial angular momentum from planetary orbits to spins, significantly changing the planetary rotation rates and axes.
Indeed, the architectures of observed exoplanetary systems imply that many of them may have undergone episodes of dynamically instability~\citep{Rasio.1996.Sci, Weidenschilling.1996.Natur, Chatterjee.2008.ApJ, Ford.2008.ApJ, Juric.2008.ApJ, Anderson.2020.MNRAS}, during which planet-planet collisions may occur and alter the planetary spins.

Our previous work has investigated the outcomes of giant-planet collisions.
In~\cite{LJR.2021.MNRAS}, we performed a suite of smoothed particle hydrodynamic simulations of close encounters between two Jupiter-like planets. 
We found that quick mergers occur when two planets physically collide, conserving more than $97\%$ of the total mass; if the incoming trajectory between the two planets has a pericenter separation less $70\%$ of the total planetary physical radius, more than $95\%$ of the planetary orbital angular momentum are inherited by the merger products as spin angular momentum.
In~\cite{LJR.2020.ApJ}, we used an ensemble of $N$-body simulations to study the giant-planet spin magnitude ($|\vec{S}|$) and obliquity ($\theta_{\rm SL}$) distributions shaped by the dynamical instability of their host planetary systems.
Assuming that two colliding planets merge into one that conserves the total mass and momentum, (as validated by the hydrodynamical simulations of~\citealt{LJR.2021.MNRAS}) we found from both simulations and analytical arguments that spin magnitudes distribution is proportional to $|\vec{S}|$ and $\text{cos}\theta_{\rm SL}$ follows a U-shape distribution.

On the other hand, the obliquity and spin magnitude caused by collisions between low-mass planets, such as super-Earths and mini-Neptunes, are less explored.
These planets are the most abundant among the detected exoplanets, and they are more susceptible to collisions than the giants.
Super-Earths and mini-Neptunes are often found in tightly-packed multi-planet systems~\citep[e.g.,][]{Lissauer.2011.ApJS, Fabrycky.2014.ApJ}.
Dynamical models indicate that many of these systems have orbital spacings almost equal to their theoretical stability thresholds, suggesting that they could be initially unstable, then evolved to their current orbital structures via collisions~\citep{Pu.2015.ApJ, Ghosh.2024.MNRAS}.
Due to the difference in their masses and physical radii, the collision-induced spin properties for low-mass planets can be different from those for gas giants. 

In this paper, we study low-mass planet obliquity and spin magnitude distributions due to collisions triggered by dynamically instability. 
Similar to~\cite{LJR.2020.ApJ}, we perform a series of $N$-body simulations of closely packed low-mass planets systems. Assuming that two colliding planets merge into a bigger one with no mass and angular momentum loss, we derive the statistics for the final planetary spin.
Although it has been noted in some hydrodynamic simulations that collisions between low-mass planets may not always lead to perfect planetary mergers~\citep{Hwang.2017.MNRAS, Hwang.2018.ApJ, Ghosh.2024.AJ}, the orbital dynamics that leads to the collisions should have the dominant effect in determining the spin distributions.
We focus on planets with masses and physical radii that resemble the detected super-Earths and mini-Neptunes, although we also conduct parameter studies using different planetary masses, radii, and initial orbital inclinations.

The rest of this paper is organized as follows. 
In Section~\ref{sec:2 results}, we present the simulation results for systems initially having two planets. 
We then study systems with three initial planets in Section~\ref{sec:Threeplanet}.
In Section~\ref{sec:conclusion}, we summarize our findings and discuss the limitations.

\section{Two-planet Cases}
\label{sec:2 results}

\subsection{Setup and Method}
\label{sec:Numericalexperiments}

We consider a planetary system consisting of a solar-type central star with mass $M_{\star}=M_{\sun}$ and physical radius $R_{\star}=R_{\sun}$, orbited by two planets of masses $m_1$ and $m_2$.
We focus on two fiducial cases: (1) super-Earth system with $m_1=m_2=3M_{\earth}$ and $R_1=R_2=1.3R_{\earth}$, and (2) mini-Neptune system with $m_1=m_2=9M_{\earth}$ and $R_1=R_2=2.4R_{\earth}$.
The planets are initially placed on compact, nearly coplanar orbits with 
\begin{align}
    a_2 - a_1 = 2.5R_{\rm H, mut},
\end{align}
where $a_1$ and $a_2$ are the orbital semi-major axes of $m_1$ and $m_2$, and
\begin{align}
\label{eq:R_Hill}
    R_{\rm H,mut} = \frac{a_1+a_2}{2}\left(\frac{m_1+m_2}{3M_\star}\right)^{1/3}
\end{align}
is the mutual Hill radius.
This configuration ensures that the mutual gravity between the planets is strong enough for the system to become dynamically unstable \citep{Gladman.1993.Icarus}.
We choose $a_1=0.1$ au in this study.

For each configuration of systems, we perform a suite of $10^4$ $N$-body simulations to study the statistical outcomes.
The initial eccentricities of the planets are randomly sampled from uniform distributions $e_{1,2}\in(0.01, 0.05)$. For initial inclinations, we assume that the probability distribution function $f(i)$ is proportional to the $\sin i$, with $i_{1,2}\in(0^\circ, i_{\rm max})$ and the fiducial value $i_{\rm max}=2^{\circ}$. 
The angular elements, including the arguments of pericenter, longitude of ascending node, and mean anomaly, are drawn uniformly from $[0, 2\pi)$.

We simulate the dynamical evolution of the system using the $N$-body code \textsc{REBOUND} \citep{Rein.2012.AandA} with the \textsc{IAS15} integrator \citep{Rein.2014.MNRAS}, which accurately resolves close encounters.
The simulations stop when one of the following occurs:
\begin{itemize}
    \item Planet-planet collision: when the separation between the two planets is less than $R_1+R_2$,
    \item Planetary ejection: when the distance of a planet from the star exceeds 100 au,
    \item Star-planet collision:  when the distance of a planet from the star is less than $R_\star+R_{1,2}$, 
    \item Time limit: when the simulation reaches a time of $3\times10^{4}$ orbital period of the inner planet. 
\end{itemize}
In practice, we find that almost all of the simulations end in planet-planet collisions, with a fraction greater than at least $95\%$. 
We will only include those that result in collisions in our analysis.

In the post-simulation analysis, we calculate the distributions of the spin and obliquity of the planet-planet collision products.
Following \cite{LJR.2020.ApJ}, we assume the two initial planets $m_1$ and $m_2$ are non-spinning--This is likely unrealistic, but our goal is to determine the baseline result for the planet's spin when initially there is none.
When they collide, the two planets will merge into a new planet in the sticky-sphere manner: the merger product will inherit the total mass, linear momentum, and angular momentum of their parents (see \citealt{LJR.2021.MNRAS}, cf. \citealt{Hwang.2018.ApJ}).

The spin angular momentum of the merger product is calculated as  
\begin{align}
\label{eq:Spin}    
    \vec{S} & = \mu \vec{r}_{\rm rel} \times \vec{v}_{\rm rel},
\end{align}
where $\mu$ is the reduced mass of the two planets, $\vec{r}_{\rm rel}$ and $\vec{v}_{\rm rel}$ are the relative position and velocity between the planets at the moment of impact \citep[see Figure 1 of][]{LJR.2020.ApJ}.
We will normalize of the spin magnitude $|\vec{S}|$ by
\begin{align}
    S_{\rm max} = \mu\sqrt{2G\left(m_1+m_2\right)\left(R_1+R_2\right)},
\end{align}
which corresponds to the magnitude of $|\vec{S}|$ when $\vec{r}_{\rm rel}$ and $\vec{v}_{\rm rel}$ are perpendicular and $|\vec{v}_{\rm rel}|$ equals the escape speed from $R_1+R_2$.

The planetary obliquity of the merger product is calculated via  
\begin{align}
\label{eq:S}    
    \cos{\theta_{\rm SL}} & = \frac{\hat{\textbf{n}}\cdot\vec{S}}{|\vec{S}|},
\end{align}
where $\hat{\textbf{n}}$ is the unit normal vector of the merger product's orbital plane.

\begin{figure*}[t]
    \centering
    \epsscale{1.15}
    \plotone{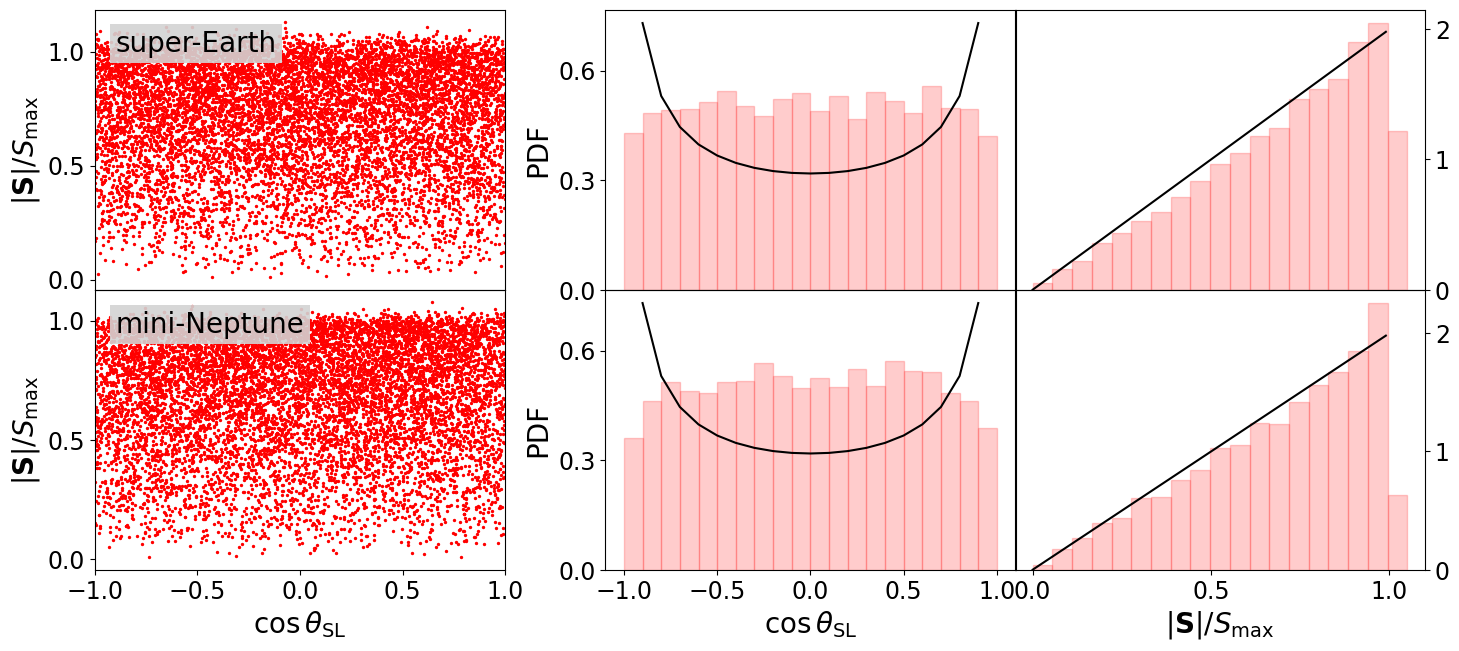}
    \caption{Spin and obliquity distributions due to planet-planet collisions in our fiducial simulations. 
    The top row is from the super-Earth simulations (i.e., with $m_1 = m_2 = 3M_{\oplus}$ and $R_1=R_2=1.3R_{\oplus}$), while the bottom row is the mini-Neptune simulations (i.e., with $m_1 = m_2 = 9M_{\oplus}$ and $R_1=R_2=2.4R_{\oplus}$). 
    The first column shows the spin magnitudes and obliquities of the merger products found in our runs. The second and third Columns show the obliquity and spin magnitude distributions of the merger products. 
    All simulations adopt $a_1=0.1$au and initial inclination $i_{1,2} \in (0^\circ, 2^\circ)$.
    The black lines are the analytical distributions from \cite{LJR.2020.ApJ}, see also Equations~\eqref{eq:f_cos} and~\eqref{eq:f_S} from Section~\ref{subsec:reason}.
    }
    \label{fig:fid}
\end{figure*}

\subsection{Fiducial Results}

We first analyze the outcomes of our fiducial simulations. 
Figure~\ref{fig:fid} shows the obliquity and spin distributions of the merger products, with the top and bottom rows showing the results for super-Earths and mini-Neptunes, respectively.
Despite the difference in planetary masses and sizes, the resulting distributions are broadly similar between the two cases.
The obliquity distributions are consistent with being nearly uniform in $\cos{\theta_{\rm SL}}$, with slight concentrations at $\cos{\theta_{\rm SL}}=0$.
The spin magnitude distribution is linear in $|\vec{S}|/S_{\rm max}$.
The black curves in the right two columns denote the analytical expectations derived by \cite{LJR.2020.ApJ} for giant-planet collisions; while the giant-planet theory agrees with the $|\vec{S}|/S_{\rm max}$ results very well, it fails to predict the $\cos{\theta_{\rm SL}}$ distribution. 
We will discuss this deviation in Section~\ref{subsec:reason}.

We note that a small number of simulations produce spin magnitudes exceeding $S_{\rm max}$. 
This occurs when the relative velocity between the two planets at contact exceeds the escape velocity.
Such cases are rare but not impossible, as the encounter between low-mass planets can be hyperbolic. 
Although it is beyond the scope of this paper, we expect these ultra-rapidly rotating merger products would spin down through mass shedding until $|\vec{S}|/S_{\rm max}\approx1$.

\subsection{Parameter Study}
\label{subsec:ps}

To explore how the physical and orbital properties of the planets affect the spin and obliquity distributions, we perform additional suites of simulations with different planetary physical radii, masses, and initial orbital inclinations.
To better highlight the effects of each parameter, we randomly sample the initial orbital inclination from $i_{1,2}\in(0^{\circ},i_{\rm max})$ but with the canonical value for $i_{\rm max}$ changed into $0.5^{\circ}$.

\begin{figure*}[t!]
    \centering
    \epsscale{0.9}
    \plotone{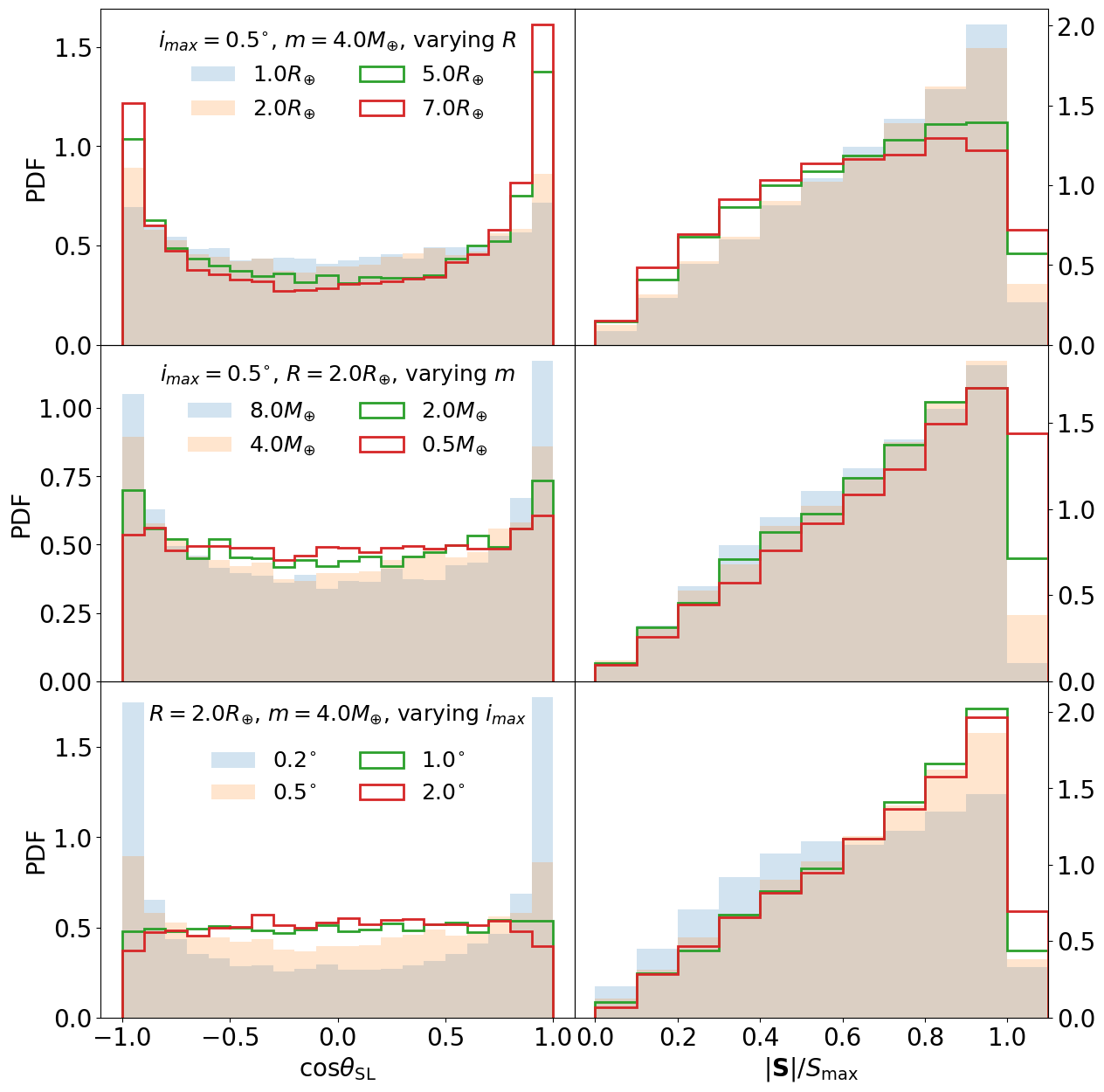}
    \caption{Parameter study for the spin and obliquity distributions of the collision products.
    {\bf Top:} for planets with $m_1 = m_2 =  4.0M_{\earth}$, $i_{\rm max}=0.5^{\circ}$,  and different physical radii.
    {\bf Middle:}: for planets with $R_1 = R_2 =  2.0R_{\earth}$, $i_{\rm max}=0.5^{\circ}$,  and different masses.
    {\bf Bottom:} for planets with $R_1 = R_2 =  2.0R_{\earth}$, $m_1 = m_2 = 4.0M_{\earth}$, and different $i_{\rm max}$.
    }
    \label{fig:ps}
\end{figure*}

The top row of Figure~\ref{fig:ps} shows the distributions of spin and obliquity when the planets initially have masses $m_{1,2}=4M_{\earth}$ but different physical radii. 
As the radius increases, the obliquity distribution becomes more polarized, with an enhanced probability at $\cos{\theta_{\rm SL}}\approx\pm1$.
The distributions of spin become uniform at high $|\vec{S}/S_{\rm max}|$ for larger $R_{1,2}$.
Both trends can be understood via a geometric argument: as their physical radii increase, the nearly coplanar planets are more likely to collide near their equators; hence, the spin angular momentum tends to be aligned or anti-aligned with the orbital plane.
Confining the impact location to the near-equator region also lowers the chance for reaching large relative angle between $\vec{r}_{\rm rel}$ and $\vec{v}_{\rm rel}$, so the probability of producing large final spin decreases.

The middle row of Figure~\ref{fig:ps} shows the effects of changing the planetary masses between $0.5M_{\earth}$ and $8M_{\earth}$.
Overall, the influence of different planetary masses is small. 
As the mass decreases, the obliquity distribution becomes less polarized, and eventually approaches a smooth and nearly uniform distribution. 

The bottom row of Figure~\ref{fig:ps} shows the important effects of the initial mutual inclination by adopting different $i_{\rm max}$.
For systems with small initial inclinations ($i_{\rm max}\leq 0.5^{\circ}$), the resulting obliquity distributions show clustering at $\cos{\theta_{\rm SL}}\approx\pm1$, while the spin distribution tends to have a flat tail at $|\vec{S}|/S_{\rm max}\sim 1$. 
For larger inclinations $(i_{\rm max}\geq1^{\circ})$, the results are similar to the fiducial, with the obliquity distribution being close to uniform and the spin magnitude distribution being linear.
This trend can be explained similarly to the radius-dependence: small inclinations restrict collisions to the equatorial regions, while larger inclinations allow impacts over a wider range of latitudes on the planetary surface.

\subsection{Analysis}
\label{subsec:reason}

\subsubsection{Analytical Results}
\label{subsec:Previousresult}
Our previous work on giant-planet collisions~\citep{LJR.2020.ApJ} predicts that the probability density distribution for $\cos\theta_{\rm SL}$ is given by
\begin{align}
\label{eq:f_cos}
    f_{\cos \theta_{\mathrm{SL}}}=\frac{1}{\pi} \frac{1}{\sqrt{1-\cos ^{2} \theta_{\mathrm{SL}}}},
\end{align}
and the distribution of $|\vec{S}|/S_{\rm max}$ is given by
\begin{align}
\label{eq:f_S}
    f_{|\vec{S}| / S_{\max }} = \frac{2|\vec{S}|}{S_{\max }}.
\end{align}
These distributions are analytically derived under the assumption that, when two planets collides, impact location is uniformly distributed in the plane perpendicular to the impact velocity (and the impact velocity lies in the original orbital plane).
\cite{LJR.2020.ApJ} show that the conditions 
\footnote{\cite{LJR.2020.ApJ} use $(a_2-a_1)/(2a_1)$ instead of $R_{\rm H,mut}/a_1$; this makes no difference for $K\sim 1$ to 3.}
\begin{align}
\frac{R_{\mathrm{p}}}{a_{1}} \ll 
\sin i \ll \frac{R_{\rm H,mut}}{a_{1}} \simeq \left(\frac{m_{1}+m_{2}}{3 M_{*}}\right)^{1 / 3}
\label{eq:limination}
\end{align}
must be satisfied for these analytical distributions to apply, where $i$ is the initial mutual inclination of the planetary orbits.
To recapitulate, satisfying the first condition $R_{\mathrm{p}}/a_{1} \ll \sin i$ allows the vertical excursion of the planets be at least their physical radius $R_{\rm p}$ as they orbit around the star.
Hence, when the planets collide, the impact location any be at any latitude on the planetary surfaces. 
This enhances the probability for the planets to reach a larger collision impact parameter, leading to the linear distribution of the spin magnitude in Eq.~\eqref{eq:f_S}.
To see the role of the second condition $\sin i \ll R_{\rm H,mut}/{a_{1}}$, let $\hat{\textbf{n}}$ be the unit normal vector of the planetary orbital plane.
When two planets enter their mutual Hill sphere, their relative velocity $\vec{v}_{\rm rel}$ has an in-plane component given by the ``Hill velocity''
\begin{align}
    v_{\parallel} \sim \sqrt{\frac{GM_{\star}}{a_1^3}}R_{\rm H,mut}
\end{align}
and a perpendicular component
\begin{align}
    v_{\perp} = \vec{v}_{\rm rel} \cdot \hat{\textbf{n}} \sim \sin{i}\sqrt{\frac{GM_{\star}}{a_1}}.
\end{align}
The relation $\sin i \ll R_{\rm H,mut}/a_1$ forces the collision to have $v_{\parallel} \gg v_{\perp}$ so that $\vec{v}_{\rm rel}$ lies in the orbital plane. The uniform distribution of $\vec{r}_{\rm rel}$ in the plane perpendicular to $\vec{v}_{\rm rel}$ then gives rise to Eqs.~\eqref{eq:f_cos} and~\eqref{eq:f_S}.

On the other hand, when the conditions
\begin{align}
    \sin i \lesssim \frac{R_{\rm p}}{a_1} \ll \frac{R_{\rm H,mut}}{a_1}
\end{align}
are satisfied, we still have $\vec{v}_{\rm rel}$ in the orbital plane, but now $\vec{r}_{\rm rel}$ also lies close to the orbital plane, so the collisions are essentially two dimensional. 
In this case, $|\vec{S}|$ distribution is uniform, i.e.,
\begin{align}
    f_{|\vec{S}|/S_{\rm max}} = \text{const},
\end{align}
while $\cos \theta_{\rm SL}$ only take two distinct values ($\pm1$).

Furthermore, when $i$ is large so that
\begin{align}
    \frac{R_{\rm p}}{a_1} \ll \frac{R_{\rm H,mut}}{a_1} \lesssim \sin i,
\end{align}
the perpendicular velocity component $v_{\perp}$ becomes comparable to the in-plane component $v_{\parallel}$.
In this case, the resulting spin vector $\vec{S}$ no longer preferentially aligns or anti-aligns with $\vec{n}$.
The probability for $\cos{\theta_{\rm SL}}=\pm1$ decreases, while other $\cos{\theta_{\rm SL}}$ values become more likely.
As a result, the obliquity distribution turns into
\begin{align}
    f_{\cos{\theta_{\rm SL}}} \simeq \text{const},
\end{align}
while the spin magnitude distribution still follows Eq.~\eqref{eq:f_S}.

\subsubsection{Analysis of the Fiducial Systems}

While Eq.~\eqref{eq:limination} is likely to be true for giant planets, it may not apply to the low-mass planets considered in this paper.
Our fiducial super-Earth systems have 
\begin{align}
    \frac{R_{\mathrm{p}}}{a_{1}} = 5.5\times10^{-4}, \  \sin{i_{\rm max}}=0.035, \ \frac{R_{\rm H,mut}}{a_{1}} \simeq 0.019,
\end{align}
while the fiducial mini-Neptunes systems have 
\begin{align}
    \frac{R_{\mathrm{p}}}{a_{1}} = 1.0\times10^{-3}, \  \sin{i_{\rm max}}=0.035, \ \frac{R_{\rm H,mut}}{a_{1}} \simeq 0.027.
\end{align}
Hence, as shown in Figure~\ref{fig:fid}, both kinds of planets satisfy the first condition $R_{\mathrm{p}}/a_{1} \ll \sin i$, so the fiducial spin magnitude distributions agree with Eq.~\eqref{eq:f_S}.
However, both cases violate the second condition $\sin i \ll {R_{\rm H,mut}}/{a_{1}}$, so the U shaped predicted by Eq.~\eqref{eq:f_cos} is washed out, and the resulting $\cos{\theta_{\rm SL}}$ distributions are nearly flat.

\begin{figure*}
    \centering
    \epsscale{0.9}
    \plotone{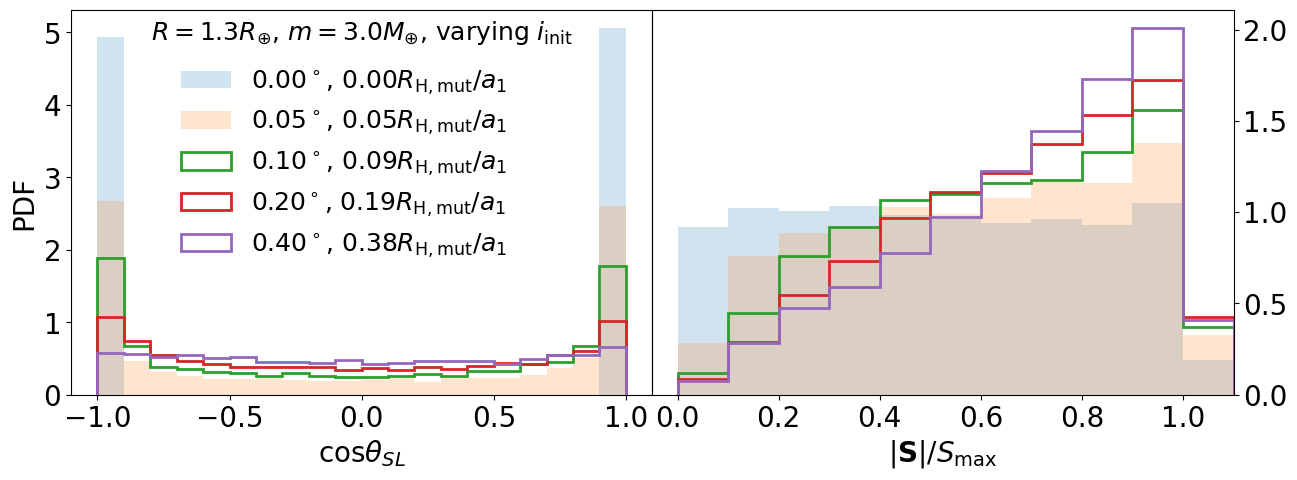}
    \caption{The obliquity and spin distribution of merger products with different fixed $i_{\rm init}$ when $R_1 = R_2 = 1.3R_{\earth}$ and $m_1 = m_2 =3M_{\earth}$.
    }
    \label{fig:fix_i}
\end{figure*}

To better understand the impact of initial inclinations, we repeat our experiments with the initial mutual inclination to be exactly $i$.  
Figure~\ref{fig:fix_i} shows the results for different $i$ values. 
It can be seen that the inclination has a great influence on the obliquity distribution. 
The results show in Figure~\ref{fig:fix_i} are consistent with the analytical expection discussed in Section~\ref{subsec:Previousresult}.

\subsubsection{Analysis of the Parameter Studies}

The trends observed in Figure~\ref{fig:ps} from the parameter studies can also be understood using our analysis in Section \ref{subsec:Previousresult}.

Increasing the planetary radius or decreasing the inclination $i_{\rm max}$ weakens the first inequality in Eq.~\eqref{eq:limination}.
Collisions then concentrate more in the equatorial region of the planets, suppressing high-latitude impacts. 
Hence, as shown in the top and the bottom rows of Figure~\ref{fig:ps}, the final distribution is more polarized in $\cos{\theta_{\rm SL}}$ and more flattened for $|\vec{S}|/S_{\rm max}$. In contrast, reducing the radius or increasing the inclination will promote high-latitude impacts.

Increasing the planetary mass slightly increase $R_{\text{H,mut}}$, so the $\cos{\theta_{\rm SL}}$ distribution becomes slightly more polarized for higher masses, as Figure~\ref{eq:R_Hill} shows.

\section{Three-planet results}
\label{sec:Threeplanet}

\subsection{Setup}

We now consider systems containing initially three planets.
As in Section \ref{sec:Numericalexperiments}, the planets are initially placed on closely packed, nearly coplanar orbits with constant mutual Hill spacing, i.e.,
\begin{align}
    \frac{a_2 - a_1}{R_{{\rm H},12}} = \frac{a_3 - a_2}{R_{{\rm H},23}} = K,
\end{align}
where $K$ is a dimensionless orbital spacing, and $R_{{\rm H},12}$ and $R_{{\rm H},23}$ are the mutual Hill radii for $m_1$-$m_2$ and $m_2$-$m_3$.
We again set $a_1=0.1\mathrm{au}$. 

Table~\ref{tab:three-planet-sim-list} lists the parameter values for the simulations in this section. 
For each configuration, we run $10^4$ simulations. 
Since there are a small number of cases where there is no collision, the sum of the percentages of each row in the table may not exactly equal to 1.

\begin{table}[]
\centering
\caption{System configurations adopted in Section~\ref{sec:Threeplanet} and the percentage of collisions between each planetary pair ($m_1 \& m_2$, $m_2 \& m_3$, $m_1 \& m_3$). 
\textbf{Upper table:} Fiducial cases. 
The super-Earth systems adopt $R_{1,2,3} = 1.3R_\oplus$ and $m_{1,2,3} = 3M_\oplus$, while the mini-Neptune systems adopt $R_{1,2,3} = 2.5R_\oplus$ and $m_{1,2,3} = 9M_\oplus$. 
Both use $i_{\max} = 2^\circ$ and $K = 3$.
\textbf{Lower table:} Parameter study cases. Unless otherwise noted, the planets are initialized with $R_{1,2,3} = 2R_\oplus$, $m_{1,2,3} = 4M_\oplus$ and $i_{\max} = 0.5^\circ$.
}
\label{tab:three-planet-sim-list}
%\vspace{-3mm}
\begin{tabular}{c|ccc}
\hline 
Fiducial Cases  & $m_1$ \& $m_2$  & $m_2$ \& $m_3$  & $m_1$ \& $m_3$  \\
\hline 
Super-Earth & 40\% & 28\% & 25\% \\
Mini-Neptune  & 37\% & 31\% & 26\%\\
\hline 
\end{tabular} \\
\begin{tabular}{c|ccc}
\hline 
Parameter Study  & $m_1$ \& $m_2$  & $m_2$ \& $m_3$  & $m_1$ \& $m_3$  \\
\hline 
$R=R_{\earth}$  & 43\% & 27\% & 23\% \\
$R=2R_{\earth}$  & 49\% & 26\% & 20\% \\
$R=5R_{\earth}$  & 60\% & 26\% & 14\% \\
$R=7R_{\earth}$  & 63\% & 25\% & 11\% \\
\hline 
$m=8M_{\earth}$  & 43\% & 30\% & 20\% \\
$m=4M_{\earth}$  & 49\% & 26\% & 20\% \\
$m=2M_{\earth}$  & 55\% & 23\% & 18\% \\
$m=0.5M_{\earth}$  & 68\% & 18\% & 11\% \\
\hline 
$i_{max} = 0.2^{\circ}$  & 54\% & 25\% & 17\% \\
$i_{max} = 0.5^{\circ}$  & 49\% & 26\% & 20\% \\
$i_{max} = 1.0^{\circ}$  & 44\% & 27\% & 23\% \\
$i_{max} = 2.0^{\circ}$  & 41\% & 28\% & 24\% \\
\hline 
K=3 & 41\% & 28\% & 24\%
\\
K=5 & 36\% & 35\% & 22\%
\\
K=7 & 33\% & 41\% & 19\%
\\
K=9 & 27\% & 49\% & 16\%
\\
\hline
\end{tabular}    
\end{table}

\begin{figure}
    \centering
    \epsscale{1.2}
    \plotone{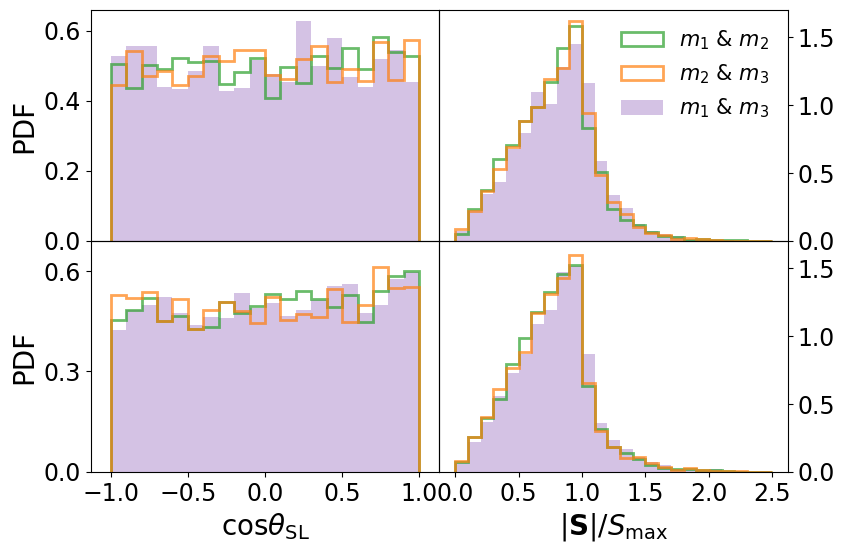}
    \caption{Obliquity and spin magnitude distribution of merger products from the fiducial three-planet simulations (see the upper panel of Table~\ref{tab:three-planet-sim-list}). 
    The top row shows the results for the super-Earth systems, and the bottom row is for the mini-Neptune systems.
    All simulations adopt $a_1=0.1$au and initial inclination $i_{1,2} \in (0^\circ, 2^\circ)$.
    The distributions shown separately for three different collision channels:
    mergers between the inner pair ($m_1$ and $m_2$, green curves), the innermost and outermost planets ($m_1$ and $m_3$,  purple histograms), and the outer pair ($m_2$ and $m_3$, orange curves).
    }
    \label{fig:3p_fid_i_2.0}
\end{figure}

\subsection{Fiducial Cases}

We first focus on two fiducial configurations:
(1) super-Earths with mass $m_{1,2,3}=3M_{\earth}$ and radius $R_{1,2,3}=1.3R_{\earth}$, and (2) three mini-Neptunes with mass $m_{1,2,3}=9M_{\earth}$ and radius $R_{1,2,3}=2.5 R_{\earth}$.
We use $i_{\rm max}=2^{\circ}$ and $K=3$ in both cases.

For three-planet systems, collisions can happen between any pair of planets. 
Table~\ref{tab:three-planet-sim-list} shows the probability for each collision channel. 
The resulting obliquity and spin distributions are shown in Figure~\ref{fig:3p_fid_i_2.0}. We see that the different collision channels lead to similar distributions.

For both super-Earth and mini-Neptune systems, the obliquity distribution of the collision products is consistent with being nearly uniform in $\cos{\theta_{\rm SL}}$, which are similar to the two-planet results shown in Section~\ref{sec:2 results}.
The distribution of $|\vec{S}|/S_{\rm max}$ is approximately linear at $|\vec{S}|<S_{\rm max}$, but also showing a noticeable tail at $|\vec{S}|>S_{\rm max}$.
It appears that three-planet systems produce more ``super-spinning'' planets with $|\vec{S}|>S_{\rm max}$ than two-planet systems;
we attribute this to the larger initial orbital spacing in three-planet systems, which can increase the impact velocity between planets.

\begin{figure}
    \centering
    \epsscale{1.2}
    \plotone{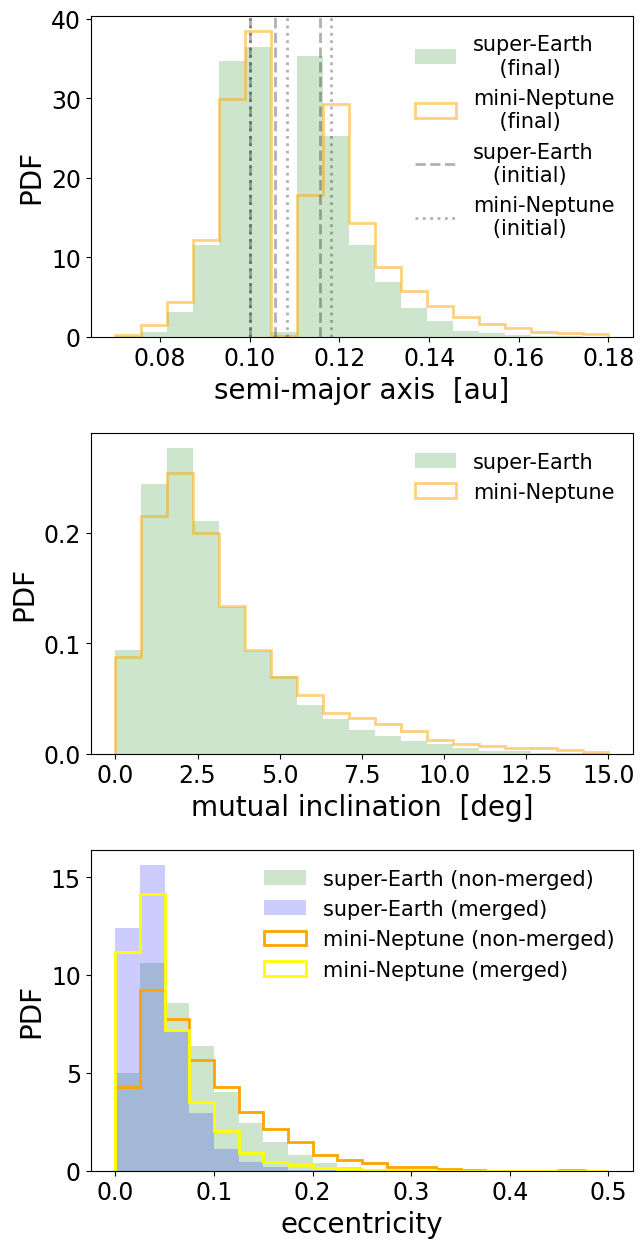}
    \caption{
    Final orbital elements in our fiducial three-planet simulations after collisions. 
    The results for the super-Earth and mini-Neptune cases are shown as filled and stepped histograms, respectively.
    {\bf Top:} Semi-major axis distributions of the planets. 
    The initial values of $a_{1,2,3}$ are marked as vertical lines. 
    {\bf Middle:} Mutual orbital inclination distributions of the planets.
    {\bf Bottom:} Eccentricity distributions of the planets, where we plot the results for the merger products and the non-collided planets separately.
    }
    \label{fig:3p_ele}
\end{figure}

For completeness, Figure~\ref{fig:3p_ele} shows the distributions of the final orbital elements, including both the merger products and the planets that do not participate collisions.
The semi-major axis distribution has two peaks, one at the initial $a_1=0.1$au and another at round $a\simeq0.12$au.
The mutual inclination peaks at $i=2^{\circ}$, but can extend up to $7^{\circ}$.
Initially uniform between $e=0.01$ and $0.05$, the eccentricity distribution also broadens to a final range between 0 and 0.2, with a peak around $0.05$. We find that the merger products generally have smaller eccentricities than the planets that do not participate in collisions.

\subsection{Parameters study}

\begin{figure*}
    \centering
    \epsscale{0.9}
    \plotone{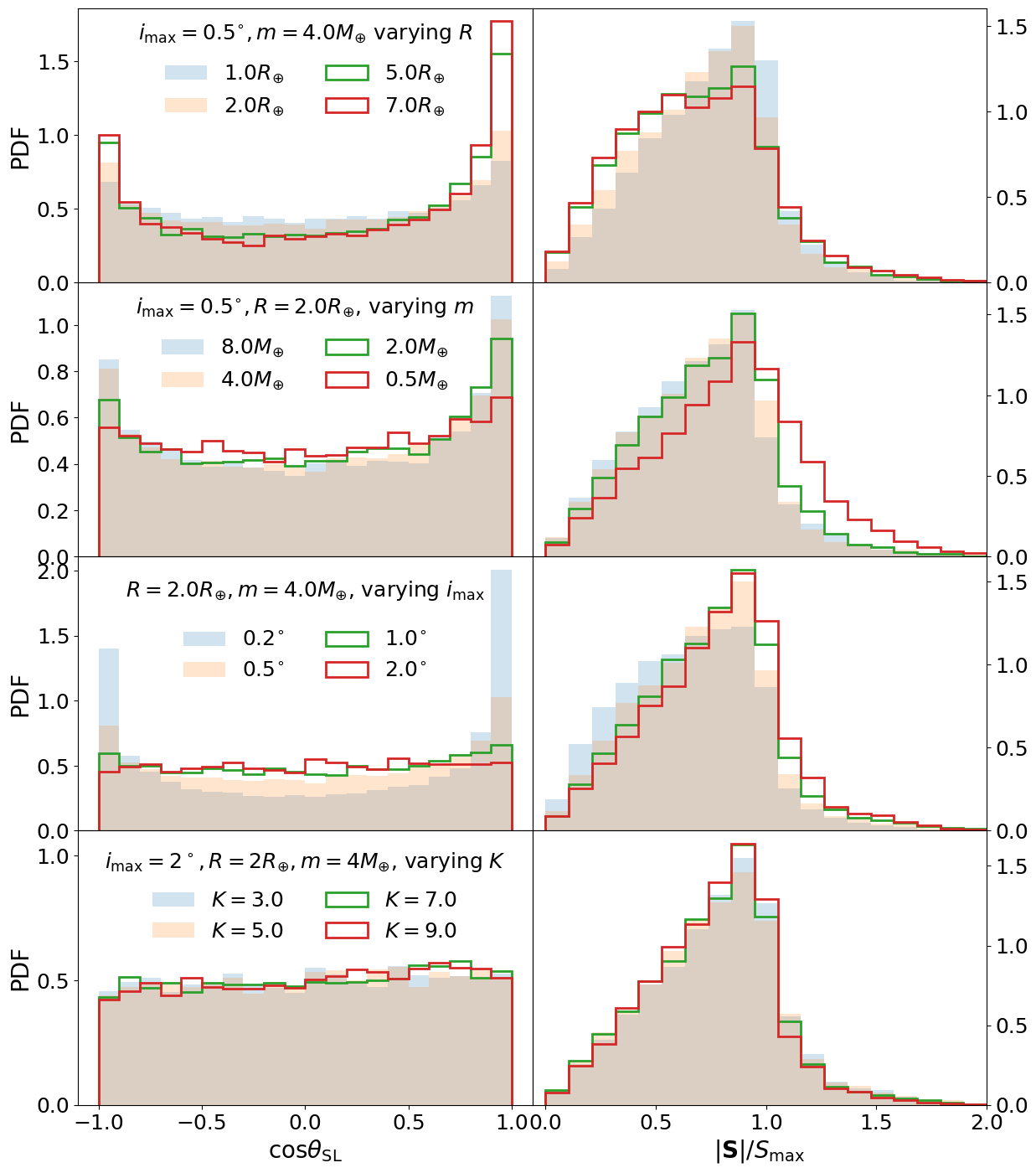}
    \caption{
    Parameter study for systems consisting three planets with $m_1 = m_2 = m_3=m$ and $R_1 = R_2 = R_3 = R$.
    {\bf Top:} simulations with $m =  4.0M_{\earth}$, $i_{\rm max}=0.5^{\circ}$, and different $R$.
    {\bf Second:} simulations with $R =    2.0R_{\earth}$, $i_{\rm max}=0.5^{\circ}$, and different $m$.
    {\bf Third:} simulations with $R = 2.0R_{\earth}$, $m_ =  4.0M_{\earth}$, and different $i_{\rm max}$. 
    {\bf Bottom:} simulations with $R = 2.0R_{\earth}$, $m = 4.0M_{\earth}$, $i_{\rm max}=2.0^{\circ}$, and different $K$.
    }
    \label{fig:3ps}
\end{figure*}

Figure~\ref{fig:3ps} compares the effects of varying the planetary radius $R$, mass $m$, orbital inclinations $i_{\rm max}$, and orbital spacing $K$.

Increasing either $R$ or $m$ drives the distribution of $\cos{\theta_{\rm SL}}$ to be more polarized towards $\pm 1$, whereas decreasing them makes the distribution more uniform.
The spin magnitude distribution at $|\vec{S}|<S_{\rm max}$ is more linear with smaller $R$ and $m$, but it flattens out at larger $|\vec{S}|/S_{\rm max}$ when $R$ or $m$ are larger.
We also find that the fraction of ``super-spinners'' ($|\vec{S}|>S_{\rm max}$) increases for smaller $R$ and $m$.
By contrast, the results for varying $i_{\rm max}$ show the opposite trend: increasing $i_{\rm max}$ flattens the $\cos{\theta_{\rm SL}}$ distribution and make the $|\vec{S}|/S_{\rm max}$ distribution more linear, while decreasing $i_{\rm max}$ enhances the polarization in obliquity and suppresses high-spin outcomes.
Overall, the effects of varying $R$, $m$ and $i_{\rm max}$ are similar to those of the two-planet systems.

Changing $K$ does not influence the distributions of obliquity and spin. However, it can be seen from Table \ref{tab:three-planet-sim-list} that increasing $K$ leads to an increase in collision percentage of $m_2$ \& $m_3$, suggesting greater chance for the two outermost planets to collide when the orbital spacing is wider.

\section{Summary and Discussion}
\label{sec:conclusion}

We have carried out a series of numerical experiments to investigate the spin magnitude and obliquity distributions of low-mass exoplanets that have gone through collisions due to dynamical instability. 

In our fiducial super-Earth ($m=3M_{\oplus}$, $R = 1.3R_{\oplus}$) and mini-Neptune ($m = 9M_{\oplus}$, $R = 2.5R_{\oplus}$) systems, the collision products follow an obliquity distribution such that $\cos{\theta_{\rm SL}}$ is nearly uniform.
The spin-magnitude distribution is approximately a linear function of $|\vec{S}|/S_{\rm max}$.

Our parameter study shows that both distributions depend on the physical and orbital parameters of the colliding planets.
Increasing planetary radii or masses, or decreasing mutual inclinations, tend to polarize the obliquity distribution toward alignment or anti-alignment (i.e., excess probability near $\cos{\theta_{\rm SL}}=\pm1$); it also tends to flatten the spin distribution at large values of $|\vec{S}|/S_{\rm max}$.
These trends in the spin and obliquity distributions can be understood analytically (See Section ~\ref{subsec:Previousresult})

Experiments with two-planet and three-planet systems produce qualitatively similar distributions, suggesting the trends reported in our results may also generalize to systems with higher planetary multiplicities.

Several limitations of our study should be noted.
For example, we use the so-called ``sticky-sphere'' prescriptions for planetary collisions; the merger products inherit the total orbital angular momentum of the two colliding planets. 
This treatment neglects several potentially important details, such as the interior structures of the planets, hydrodynamics of collisions, and planetary mass loss during impacts.
Hydrodynamic simulations have begun to explore some of these effects ~\citep[e.g.,][]{LJR.2021.MNRAS, Ghosh.2024.AJ}.
To incorporate these collisional effects in the obliquity predictions, future studies may combine $N$-body simulations with fitting formulae derived from these hydrodynamic results~\citep[e.g.,][]{Ghosh.2024.AJ}.
Another caveat is that we have restricted our analysis to the systems experiencing at most one collisions.
In reality, closely packed planets systems may undergo multiple episodes of planet-planet collisions, which could further reshape the final spin magnitude and obliquity distributions.

Despite these possible caveats, our results demonstrate that planet-planet collisions plays a fundamental role in shaping the rotational properties of low-mass planets in closely packed systems.
Planets that have experienced collisions are expected to be rapidly spinning, with their obliquities follow an isotropic distribution.
These collision-induced spin and obliquity provide a broad range of initial conditions for various kinds of subsequent evolutionary processes, such as tidal damping of obliquity and secular interactions through spin–orbit coupling~\citep{Su.2022.MNRAS, Su.2022.MNRAS.513}. 
As observations begin to constrain the spin properties of exoplanets, our results help to bring insights into their dynamical histories.

\begin{acknowledgments}

This work is supported in part by T. D. Lee Institute and computation performed at the astro cluster of TDLI.
Jiaru Li is supported by a CIERA Postdoctoral Fellowship.
Part of this work used computing resources provided by Northwestern University and the Center for Interdisciplinary Exploration and Research in Astrophysics (CIERA). 
This research was supported in part through the computational resources and staff contributions provided for the Quest high performance computing facility at Northwestern University which is jointly supported by the Office of the Provost, the Office for Research, and Northwestern University Information Technology.
\end{acknowledgments}

\end{document}